\documentclass[a4paper,dvips,12pt]{article}
\usepackage{amsmath,amssymb,exscale}   
\usepackage{array,multicol}
\usepackage{afterpage,float,flafter}   
\usepackage{epsfig,rotating,pifont,fancybox}   
\usepackage{cite}
\usepackage{axodraw}
\makeatletter   
\newcommand{\bea}{\begin{eqnarray}}   
\newcommand{\eea}{\end{eqnarray}} 
\newcommand{\be}{\begin{equation}}   
\newcommand{\ee}{\end{equation}}

\def\ctu(#1,#2){
\BCirc(#1,#2)4
\put(0,0){
}
\put(0,0){
}
}
\def\ctq(#1,#2){
\BCirc(#1,#2)4
\put(0,0){
}
\put(0,0){
}
}

\def\a{\alpha}

\def\g{\gamma}

\def\d{\delta}

\def\m{\mu}
\def\n{\nu}

\def\tr{{\rm tr}}

\title{   
\vspace*{-0.8cm}   
\begin{flushright}   
\normalsize{      
IEM-FT/227-02\\
\texttt{hep-ph/0210134}}\\ 
\end{flushright}    
\vspace{1cm}
\Large{\sc Radiative brane-mass terms in $D>5$ orbifold gauge theories~\footnote{Work 
supported in part by CICYT, Spain, under contracts FPA2001-1806 and FPA2002-00748,
and by EU under contracts HPRN-CT-2000-00152 and HPRN-CT-2000-00148.}}
\vspace*{.5cm}
\author{\large
{\sc G.~v.~Gersdorff, N.~Irges  and M.~Quir{\'o}s}\\ \\
\emph{Instituto de Estructura de la Materia (CSIC), Serrano 123,}\\
\emph{E-28006-Madrid, Spain.}}}
\date{}   
\begin{document}
\maketitle
\thispagestyle{empty}
\vspace*{.5cm}

\begin{abstract}\noindent
A gauge theory with gauge group $\mathcal{G}$ defined in $D>4$
space-time dimensions can be broken to a subgroup $\mathcal{H}$ on
four dimensional fixed point branes, when compactified on an orbifold.
Mass terms for extra dimensional components of gauge fields $A_i$
(brane scalars) might acquire (when allowed by the brane symmetries)
quadratically divergent radiative masses and thus jeopardize the
stability of the four-dimensional theory. We have analyzed
$\mathbb{Z}_2$ compactifications and identified the brane symmetries
remnants of the higher dimensional gauge invariance. No mass term is
allowed for $D=5$ while for $D>5$ a tadpole $\propto F_{ij}^\alpha$
can appear when there are $U_\alpha(1)$ factors in $\mathcal{H}$. A
detailed calculation is done for the $D=6$ case and it is established
that the tadpole is related, although does not coincide, with the
$U_\alpha(1)$ anomaly induced on the brane by the bulk fermions. In
particular, no tadpole is generated from gauge bosons or fermions in
real representations.

\end{abstract}
\vspace{3.cm}   
   
\begin{flushleft}   
October 2002 \\   
\end{flushleft}
\newpage

An important issue in any model of particle physics is how the
mechanism of electroweak symmetry breaking is realized. In the
Standard Model this is achieved by introducing the Higgs field.
However, this phenomenologically well motivated mechanism comes with
an undesired effect. The Higgs mass should be of the order of the
electroweak symmetry breaking scale ($\sim 10^2$ GeV), which is
unnaturally small compared to the ultraviolet (UV) cut-off of the
Standard Model ($\sim 10^{19}$ GeV). In addition 
the hierarchy of scales is destroyed by radiative corrections and fine
tuning is required to keep the Higgs light. This is commonly known as
the hierarchy problem and one might wish to extend the Standard Model
by supersymmetry to soften the UV sensitivity of the scalar sector.

However, introducing extra space-time dimensions opens new ways to
solve the hierarchy problem. On the one hand, in the presence of
transverse (as large as submillimeter) dimensions where only gravity
propagates the scale of quantum gravity (string scale) can be lowered
from the Planck scale to the TeV
range~\cite{Lykken:1996fj,Arkani-Hamed:1998rs}, thus alleviating the
problem of quadratic divergences. On the other hand, in the presence
of (longitudinal) TeV extra dimensions~\cite{Antoniadis:1990ew} it is
not necessary to introduce any fundamental scalars at all, but instead
one could use the fact that the extra dimensional components of gauge
bosons are scalars under the four dimensional (4D) Lorentz symmetry
and transform non-trivially under the gauge symmetry they generate in
the higher dimensional theory. These scalars can be used to break
electroweak symmetry spontaneously
\cite{Hosotani:1983xw,Randjbar-Daemi:1982hi,
Hatanaka:1999sx,Antoniadis:2000tq,ABQ}.
One might then conclude that higher dimensional gauge invariance
protects those scalars from being sensitive to the UV physics.

Orbifolds~\cite{Dixon:jw} play a prominent role in theories with extra
dimensions due to their property to create chirality in the massless
sector, an indispensable property in any phenomenologically relevant
theory.  Another interesting feature of orbifolds is their ability to
break symmetries, in particular gauge and supersymmetry. While local
symmetries remain intact in the bulk by an appropriate choice of
parities for the transformation parameters, they are in general broken
to smaller subgroups on the boundaries (the fixed points of the
orbifold symmetry).  As in any quantum field theory, in the effective
action we must allow for all operators consistent with the
symmetries. Allowed operators not present at tree level will be
generated by radiative
corrections~\cite{Georgi:2000ks,Goldberger:2001tn,
vonGersdorff:2002as,Cheng:2002iz}.

The orbifold breaking of the bulk gauge symmetry proceeds by
projecting out some fields, i.e. only a subset of the 4D gauge bosons
$A_\mu$ and the 4D scalars $A_i$ ($i=5,\cdots ,D$) will be
non-vanishing at the boundaries.  While these $A_\mu$ generate the
unbroken gauge group $\cal H$, the $A_i$ transform in some
representation of $\cal H$.  It is then necessary to determine how the
symmetries restrict possible brane localized operators of those
fields, especially possible mass terms for the
scalars~\cite{vonGersdorff:2002as}. Would $\mathcal{H}$ be the only
symmetry left on the brane, mass terms for $A_i$ would be perfectly
allowed leading to a quadratic sensitivity to the UV cut-off.

In this letter we demonstrate that the remnant symmetry on the brane
is larger than the $\cal H$ gauge symmetry left over from the
bulk. This provides a further restriction on the possible brane terms.
We find that brane mass terms for scalars can only occur in $D\geq6$
and only for $U(1)$ factors in $\cal H$ that were not already
present in the bulk gauge group $\cal G$. These brane mass terms are
radiatively generated by bulk fermions.

We will consider a gauge theory (gauge group $\mathcal{G}$) coupled to
fermions in $D>4$ dimensional space-time parametrized by coordinates
$x^M=x^{\m}, y^i$ where $\m=0,1,2,3$ and $i=5,\cdots ,D$. The bulk
Lagrangian is
\begin{equation} {\cal L}_{D}=-\frac{1}{4}
{F}_{MN}^A{F}^{AMN}+i{\overline \Psi}
\g^MD_M{\Psi},\label{5daction}
\end{equation}
where $F_{MN}^A=\partial_{M}A_N^A-\partial_{N}A_M^A
-g\,f^{ABC}A_M^BA_N^C$ with the indices $A,B,C$ running over the
adjoint representation of ${\cal G}$ and $f^{ABC}$ being the ${\cal G}$
structure constants.  The local symmetry of (\ref{5daction}) is
the invariance under the (infinitesimal) gauge transformations
\begin{equation} 
\d_{\cal G} A_M^{A}=\frac{1}{g}D_M^{AB} \xi ^B=
\frac{1}{g}\partial_M\xi^A - f^{ABC}\xi^BA_M^C. 
\label{gaugetr}\end{equation}
We now compactify the $p\equiv D-4$ extra dimensions on the
$T^{p}/\mathbb{Z}_2$ orbifold with all the radii of the torus equal to
$R$~\footnote{From now on we will work in units where $R\equiv
1$. Restoring the $R$ dependence as well as introducing different
radii $R^i$ for different dimensions should be straightforward.} and
with the $\mathbb{Z}_2$ action defined as $y^{i}\rightarrow - y^{i}.$

In the compactified theory the surviving gauge symmetry on the 
boundaries of the orbifold is a subgroup ${\cal H}$ of ${\cal G}$,
according to the action of $\mathbb{Z}_2$ on the gauge fields
\begin{equation}
A(x^{\m},-y^i)={\cal P}_{\cal A} A(x^{\m},y^i),\hskip 1cm
{\cal P}_{\cal A}=\Lambda\otimes {\cal P}_1\label{Aproj}.
\end{equation}
Here ${\cal P}_1$ acts on the vector indices and it is the diagonal
matrix with eigenvalues $\a_{\m}=+1, \a_i=-1$. $\Lambda$ acts on the
gauge indices and can also be taken diagonal. Its eigenvalues
$\eta^A=\pm1$ then define the breaking pattern. We split the bulk
gauge index as $A=a, {\hat a}$ corresponding to the unbroken
($\eta^a=+1$) and the broken generators ($\eta^{\hat a}=-1$)
respectively.  The nonzero fields on the brane are the even fields,
namely $A_{\m}^{a}$ and $A_i^{\hat a}$, while $A_{\m}^{\hat a}$ and
$A_i^{a}$ are odd and thus vanish on the brane.  The orbifold
consistency constraint on the structure constants comes essentially
from the invariance of (\ref{5daction}) and it provides the
automorphism condition~\cite{Hebecker:2001jb}
\begin{equation} 
\eta^{A}\eta^{B}\eta^{C}=1,\qquad\text{for }f^{ABC}\neq 0. 
\label{auto1}
\end{equation}
Finally, in the gauge sector, the Faddeev-Popov ghosts $c$ transform as
the $\mu$-components of the gauge fields, and for them the parity
action is $\mathcal{P}_c=\Lambda$.

There are restrictions on the fermion representations as well.  In
even dimensions the bulk fermion representation has to be chosen
anomaly free. Furthermore, for any number of extra dimensions, the
resulting four dimensional massless fermion spectrum must also be
anomaly free. In addition, there are orbifold consistency conditions
analogous to (\ref{auto1}).  The $\mathbb{Z}_2$ action on the fermions
is
\begin{equation} 
\Psi(x^{\m},-y^i)={\cal P}_{\Psi}\Psi(x^{\m},y^i),\hskip 1cm
{\cal P}_{\Psi}=\lambda \otimes {\cal P}_{\frac{1}{2}} 
\label{pfermionapp}
\end{equation}
where $\lambda$ is a matrix acting on the representation indices.  The
constraint comes from the requirement that the coupling
$iA_{M}^A{\overline \Psi}\g^{M}T^A{\Psi}$ is $\mathbb{Z}_2$
invariant. One obtains~\cite{vonGersdorff:2002as} for any number of
dimensions~\footnote{Note that conditions (\ref{auto2}) determine
$\lambda$ up to a sign.}
\begin{equation} 
[\lambda,T^a]=0\hskip 1cm \{\lambda,T^{\hat a}\}=0 \label{auto2}. 
\end{equation} 
${\cal P}_\frac{1}{2}$ is the orbifold action on the spinor indices
and will be given explicitly later on.

The non-vanishing fields on the branes are of the general form
\begin{equation} 
\prod_{i=5}^{D} \partial _{i}^{n_i}
\left.\Phi \right|_{\rm brane}\equiv \partial^n \Phi\label{Phi}
\end{equation}
where $n\equiv \sum_i n_i$ is even (odd) for even (odd) fields.
Similarly, the gauge parameters $\xi^{a}$ are even fields and
$\xi^{\hat a}$ are odd. They couple to the branes according to
(\ref{Phi}).

The effective four dimensional Lagrangian can be written as
\begin{equation}
{\cal L}_{4}^{eff}=\int d^{p}y\bigl[{\cal L}_{D}+{\cal L}_{4}^{brane}
\prod_i\left\{\d({y^i})+\d(y^i-\pi)\right\}\bigr]
\label{lag}
\end{equation}
where ${\cal L}_{D}$ is given by (\ref{5daction}) and ${\cal
L}_{4}^{brane}$ should be the most general Lagrangian consistent with
the symmetries. The latter can be nothing but the original bulk
symmetry (\ref{gaugetr}) modded out by the orbifold action and
subsequently evaluated at the location of the brane. Let us call the
transformation resulting from this operation $\d_{\xi}$.  Applying
this rule to (\ref{gaugetr}) acting on the massless even fields, one
obtains the transformations
\begin{equation}
\d_{\xi} (A_{\m}^a) = \frac{1}{g}\partial _{\m}\xi ^a - f^{abc}\xi^bA_{\m}^c,
\label{trans1}
\end{equation}
\begin{equation}
\d_{\xi} (A_{i}^{\hat a}) = \frac{1}{g}\partial _i\xi ^{\hat a} - 
f^{{\hat a}{b}{\hat c}}\xi^bA_{i}^{\hat c}
\label{trans2}.
\end{equation}
In the above equations and in what follows, all fields should be
interpreted as coupled to the brane in (\ref{lag}) according to
(\ref{Phi}).

The brane symmetry is however much larger than the transformations
(\ref{trans1}) and (\ref{trans2}).  In fact, there is an infinite
number of non-zero independent fields on the brane, i.e.
$\partial^{2k}\{ A_{\m}^a, A_i^{\hat a} \}$ and
$\partial^{2k+1}\{A_{\m}^{\hat a}, A_i^{a}\}$, and an infinite number
of corresponding transformation parameters $\{\partial ^{2k}\xi^a\}$
and $\{\partial^{2k+1}\xi^{\hat a}\}$ induced by the bulk.  Using
(\ref{gaugetr}), one can derive the transformation of any non-zero
brane field. We show explicitly only the first two at the next level:
\begin{equation}
\d_{\xi} (\partial_j A_{i}^{a}) =\frac{1}{g}
 \partial _{j} (\partial _{i}\xi ^{a})
-\, f^{{a}{\hat b}{\hat c}}(\partial _j\xi^{\hat b})A_{i}^{\hat c}
-\, f^{{a}{b}{c}}\xi^{b}(\partial _jA_{i}^{c}),
\label{trans3}
\end{equation}
\begin{equation}
\d_{\xi} (\partial_i A_{\m}^{\hat a})=\frac{1}{g}
\partial _{\m}(\partial _{i}\xi ^{\hat a})
-f^{{\hat a}{\hat b}{c}}(\partial _i\xi^{\hat b})A_{\m}^{c}
-f^{{\hat a}{b}{\hat c}}\xi^{b}(\partial _iA_{\m}^{\hat c})
\label{trans4}.
\end{equation}
It is convenient to separate the above transformations into two
different classes:
\begin{equation}
\d_{\xi}=\d_{\cal H}+\d_{\cal K}\hskip .5cm {\rm with}\hskip .5cm
\d_{\cal H}=\{ \xi^a \}, \hskip .3cm
\d_{\cal K}=\{\partial ^{2k}\xi^a, \partial ^{2k+1}\xi^{\hat a} \}
\label{split}.
\end{equation}
This is a natural separation because $\d_{\cal H}$ is the surviving
gauge transformation on the brane reflecting its $\cal H$ gauge
invariance. One can see immediately by inspection of
Eqs.~(\ref{trans1})$-$(\ref{trans4}) that $A_\mu^a$ are the gauge
bosons of $\cal H$ while all other fields transform homogeneously in
either the adjoint of $\cal H$, $(T^a)_{bc}=if^{abc}$, or in the
representation spanned by $(T^a)_{\hat b\hat c}=if^{ a\hat b\hat
c}$~\footnote{As a simple example consider the breaking
$SU(3)\rightarrow SU(2)\otimes U(1)$. The adjoint of $SU(3)$,
$f^{ABC}={\bf 8}$ then splits into the $SU(2)$ representations
$f^{abc}={\bf 3}\oplus {\bf 1}$ ($\cal H$ is not simple and hence its
adjoint is reducible) and $f^{a\hat b\hat c}={\bf 2}\oplus {\bf 2}$.
}. The rest of the transformations is a set of local (but not gauge)
transformations which we named $\d_{\cal K}$.

Once the symmetries under which the brane action should be invariant
are known, one can start constructing the allowed terms by these
symmetries. A useful guiding principle in this task is the gauge
symmetry $\cal H$. We know that it is a necessary condition that the
building blocks should be $\cal H$$-$covariant combinations of the fields
since this (and only this) can ensure that the square of these
covariant objects are $\d_{\cal H}$$-$invariant. Given a set of $\cal
H$$-$covariant objects, invariance under $\d_{\cal K}$ is a sufficient
condition for their square to be invariant under both $\d_{\cal H}$ and
$\d_{\cal K}$ and therefore to be an allowed terms in the effective
action.  The reason for which we required $\cal K$$-$invariance is
because there is no notion of $\cal K$$-$covariance, since $\cal K$ is
not a gauge symmetry.  Thus, even though at this point we have not
proved that $\cal K$$-$invariance is not only a sufficient but also a
necessary condition, we will enforce it.
 
A simple and very important example is the field $A_i^{\hat a}$.  By
looking at (\ref{trans2}) one can see that this field is indeed
$\delta_{\cal H}-$covariant but not $\delta_{\cal K}-$invariant. A
naive interpretation would then be that an explicit brane mass term as
$(A_i^{\hat a}M_{{\hat a}{\hat b}}A_j^{\hat b})$ is forbidden in the
four dimensional effective action. However, as we will see below,
under particular circumstances such a term can be part of a
$\delta_{\cal H}-$ and $\delta_{\cal K}-$invariant term in the
Lagrangian in which case such a term can be generated radiatively.

The terms which are at the same time $\cal H$$-$covariant and 
$\cal K$$-$invariant are easily found from the transformation properties:
\begin{equation}
\d_{\cal H}F_{\m\n}^{a}=-f^{{a}{b}{c}}\xi^{b}F_{\m\n}^{c},\hskip 1cm
\d_{\cal K}F_{\m\n}^{a}=0\label{cov1}
\end{equation}
\begin{equation}
\d_{\cal H}F_{i \m}^{\hat a}=-f^{{\hat a}{b}{\hat c}}
\xi^{b}F_{i \m}^{\hat c},\hskip 1cm
\d_{\cal K}F_{i \m}^{\hat a}=0\label{cov2}
\end{equation}
\begin{equation}
\d_{\cal H}F_{ij}^{a}=-f^{{a}{b}{c}}\xi^{b}F_{ij}^{c},\hskip 1cm
\d_{\cal K}F_{ij}^{a}=0\label{cov3}.
\end{equation}
Note the different structure of $F^a_{\mu\nu}\equiv\partial_\mu
A^a_\nu-\partial_\nu A^a_\mu-gf^{abc}A^b_\mu A^c_\nu$ and 
$F^a_{ij}\equiv\partial_i
A^a_j-\partial_j A^a_i-gf^{a\hat b\hat c}A^{\hat b}_i A^{\hat c}_j$
in the nonlinear terms.
Further terms
could be constructed from covariant derivatives of these operators.
At the renormalizable level the following terms can appear in the
Lagrangian:
\begin{equation} 
{\cal L}_{4}^{brane}= -\frac{1}{4}\mathcal{Z}_{ab}F_{\m\n}^aF^{b\,\m\n}
 -\frac{1}{4}\mathcal{Z}_{\hat a \hat c}^{ij}F_{i\m}^{\hat a}F_j^{\hat c\, \mu}
 -\frac{1}{4}\mathcal{Z}_{ab}^{ijkl}F_{ij}^{a}F_{kl}^{b}+
\mathcal{Z}_\alpha^{ij}F_{ij}^\alpha+\mathcal{Z}_\alpha^{klij}
D_k^{\alpha A}D_l^{AB}F_{ij}^B.
\label{renormL}
\end{equation}
where the $\mathcal Z$ tensors in extra-dimensional indices must be
proportional to either the torus metric $g^{ij}$ or to possible
invariant tensors under the symmetry group of the torus.  We
differentiate in the last two terms of (\ref{renormL}) possible $U(1)$
factors of ${\cal H}$ from the remaining semi-simple part and denote
these $U(1)$ generators by ${T^\a}$. In fact Eq.~(\ref{cov3}) implies
that the field strenght of a $U(1)$ gauge field is invariant by itself
allowing for the term~\footnote{Notice that unbroken $U(1)$ factors in
$\mathcal{G}$ do not give rise in (\ref{Fterm}) to bilinear terms in
even fields.}
\begin{equation}
\label{Fterm}
F^\a_{ij}=2\,\partial_{[i}^{\phantom{a}} A_{j]}^\a
-g f^{\alpha \hat b \hat c}A_i^{\hat b}A_j^{\hat c}.
\end{equation}
that can give rise to a quadratic renormalization. In a similar way,
the term $D_k^{\alpha A}D_l^{AB}F_{ij}^B$ is invariant allowing for
the last term in (\ref{renormL}). It is dimension four and gives rise
to a logarithmic renormalization, as we will see.

One might think that the term tr$(\lambda_RT_R^a) F_{ij}^a$, where
$\lambda_R$ satisfies Eqs.~(\ref{auto2}) and the index $R$ denotes
some arbitrary irreducible representation, would give a further
invariant linear in $F_{ij}$\footnote{We thank C.~Cs\'aki for pointing
this out to us and thus making us aware of the possibility of having
terms linear in $F_{ij}$ on the brane.}. However, for $T^a_R$
belonging to a simple factor of ${\cal H}$, $\lambda_R$ must act as
the identity in this subspace by Eqs.~(\ref{auto2}) and Schur's Lemma,
so the trace vanishes. Only $U(1)$ factors will thus contribute to the
trace and we do not get any new invariant. We conclude that the terms
$F^\a_{ij}$ are the most general linear terms.

We will be concerned mainly with the appearance of scalar mass terms
in ${\cal L}_{4}^{brane}$. For a general unbroken gauge group
$\mathcal{H}$ the most general renormalizable Lagrangian allowed by
the symmetries of the theory contains the terms in (\ref{renormL}).
The first term in (\ref{renormL}) corresponds to kinetic terms for the
four dimensional gauge bosons, the second one corresponds to kinetic
terms for the even scalars (plus some interactions), while the third
term contains brane mass terms for the odd scalars.  One consequence
of the appearance of brane mass terms in this particular way is that
their renormalization is expected to be governed by the (wave
function) renormalization of $F^2$, which does not contain quadratic
divergences. They are expected to pick up only logarithmically
divergent renormalization effects.  Brane mass terms for even scalars
can appear in ${\cal L}_{4}^{brane}$ in the case where there are
$U(1)$ group factors in $\mathcal{H}$ corresponding to unbroken
generators $T^\a$. Under this circumstance we have seen that the
operator (\ref{Fterm}) is allowed by all symmetries on the brane and
we expect that both a tadpole for the derivative of odd fields,
$\partial_iA^\a_j$, and a mass term for the even fields,
$f^{\a\hat b \hat c}A_i^{\hat b} A_j^{\hat c}$, will be
generated on the brane by bulk radiative corrections. Moreover, since
these operators have dimension two, we expect that their respective
renormalizations will lead to quadratic divergences, making the theory
ultraviolet sensitive.

We would like to confirm by explicit calculation that the
allowed terms are indeed generated radiatively on the brane. In
particular, mass terms for brane scalars (extra dimensional components
of gauge bosons) are contained in the third term of (\ref{renormL})
for the odd scalars $A_i^a$, and in (\ref{Fterm}) for the even scalars
$A_i^{\hat a}$ when there are $U(1)$ group factors in $\mathcal{H}$.
In all cases they arise from effective operators proportional to
$F_{ij}$. An important special case is $D=5$, i.e. a five dimensional
gauge theory compactified on $S^1/\mathbb{Z}_2$. In this case the term
$F_{ij}$ does not exist and therefore we do not expect any type of
brane mass terms to appear in ${\cal L}_{4}^{brane}$. This result has
been confirmed by explicit one loop calculation in
Ref.~\cite{vonGersdorff:2002as}. However for $D>5$ $F_{ij}$ does exist and we expect,
from the previous symmetry arguments, the corresponding mass terms to
be generated on the brane by radiative corrections. The rest of this
letter will be devoted to an explicit calculation of these mass terms
in a $D=6$ model compactified on the orbifold $T^2/\mathbb{Z}_2$.

In $D=6$ the Clifford algebra is spanned by eight dimensional matrices
$\Gamma_M=\Gamma_\mu,\Gamma_i$ safisfying
$\{\Gamma_M,\Gamma_N\}=2g_{MN}$. For an appropriate choice of the
representation of the $\Gamma^M$, Dirac spinors in six dimensions are
of the form $\Psi=\left(\Psi_1, \Psi_2\right)^T$ where $\Psi_{1,2}$
are Dirac spinors in the four dimensional sense. We can define the six
dimensional Weyl projector leading to the corresponding six
dimensional chirality such that $\Gamma_7\Psi_{\pm}=\pm \Psi_{\pm}$
where $\Psi_{\pm}$ are six dimensional chiral spinors. The chiral
fermions $\Psi_{\pm}$ contain the degrees of freedom of a four
dimensional Dirac spinor. A theory with six dimensional chiral
fermions is not free from six dimensional anomalies, generated by box
diagrams, and anomaly freedom should be enforced by appropriately
restricting the fermion content~\footnote{Of course a theory with six
dimensional Dirac fermions does not have six dimensional
anomalies.}. After orbifolding on $T^2/\mathbb{Z}_2$, half of the
degrees of freedom of the chiral fermions $\Psi_{\pm}$ is odd and the
zero mode sector becomes chiral from the four dimensional point of
view; four dimensional anomaly freedom for zero modes is a further
requirement of any consistent theory.

Compactification of a six dimensional theory to four dimensions is
done by means of usual techniques. The Kaluza-Klein (KK) number is now
a two dimensional vector denoted by $\vec{m}$. Generalizing the
techniques of \cite{Georgi:2000ks,vonGersdorff:2002as}
to six dimensions is easy, the only missing ingredient being the
orbifold action on the fermions $\mathcal{P}_{\frac{1}{2}}$ in
(\ref{pfermionapp}). 
Requiring that 
\begin{equation}
\mathcal{P}_{\frac{1}{2}}^{2}=1,\qquad
\mathcal{P}_{\frac{1}{2}}\Sigma_{MN}\mathcal{P}_{\frac{1}{2}}=(\mathcal{P}_1)^R_M(\mathcal{P}_1)^S_N\Sigma_{RS}
\end{equation}
where $\Sigma_{MN}=\frac{i}{4}[\Gamma_M,\Gamma_N]$, we can identify
two possible solutions:
\begin{equation}
\mathcal{P}_{\frac{1}{2}}=i\Gamma_5\Gamma_6,\qquad
\mathcal{P}'_{\frac{1}{2}}=i\Gamma_5\Gamma_6\Gamma_7.
\label{p12}
\end{equation}
Both projections differ in their action on possible discrete
space-reflection symmetries, which might be broken by the orbifold or not.
The situation is summarized in Table~\ref{tabla}.

%
\begin{table}[h]
\begin{center}
\begin{tabular}{|c|c|c|c|c|}
\hline
&$\vec{x},y^5,y^6$ (6D parity)&$\vec{x}$ (4D parity) &$y^5$ & $y^6$ \\
\hline
$\mathcal{P}_{\frac{1}{2}}$&conserved&conserved&broken&broken\\
$\mathcal{P}'_{\frac{1}{2}}$&broken&broken&conserved&conserved\\
\hline
\end{tabular}
\end{center}
\caption{\em Discrete Lorentz symmetries broken/conserved by the
orbifold. The corresponding reflected coordinates are indicated.}
\label{tabla}
\end{table}
%
The main difference is that starting from a 6D Dirac spinor, using
$\mathcal{P}'_{\frac{1}{2}}$ the massless spectrum contains two 4D Weyl
spinors of the same chirality while with $\mathcal{P}_{\frac{1}{2}}$
it contains a 4D Dirac spinor. This is consistent with the fact that
$\mathcal{P}'_{\frac{1}{2}}$ breaks 4D parity while
$\mathcal{P}_{\frac{1}{2}}$ conserves it.  We would like to stress
that the distinction is completely irrelevant in case the discrete
symmetries are broken in the first place, as is the case when dealing
with 6D Weyl fermions. We can obtain the same massless field content
with $\mathcal{P}_{\frac{1}{2}}$ and $\mathcal{P}'_{\frac{1}{2}}$
since we are now allowed to choose different $\lambda$ for different
chiralities~\footnote{\label{bla}Indeed,
$\lambda_i\otimes\mathcal{P}_{\frac{1}{2}}$ produces the same zero
mode spectrum as $\lambda'_i\otimes\mathcal{P}'_{\frac{1}{2}}$ with
$\lambda'_i=\varepsilon_i\lambda_i$, $\varepsilon_i$ being the 6D
chirality of the fermions species $\psi_i$.
}. Without loss of generality we will choose
$\mathcal{P}_{\frac{1}{2}}$ for 6D Weyl fermions.

The propagator of the $\vec{m}$-mode of an arbitrary field $\Phi$ (a
gauge boson $A_M$, a ghost field $c$ or a fermion $\Psi$) in the
six dimensional space compactified on the orbifold $T^2/\mathbb{Z}_2$
can be written as
\begin{equation}
\left<{\Phi}^{{\vec m}'}{\overline \Phi}^{\vec m}\right>=
\frac{1}{2} \left(\delta_{{\vec m}'-\vec m}+ {\cal
P}_{\Phi}\delta_{{\vec m}'+{\vec
m}}\right)G^{(\Phi)}(p_\m,p_i).
\label{propagator}
\end{equation}
where $G^{(\Phi)}(p_\m,p_i)$ is the propagator of the corresponding
field in flat six dimensional space and ${\cal P}_{\Phi}$ the parity as
defined in (\ref{Aproj}) and (\ref{pfermionapp}).

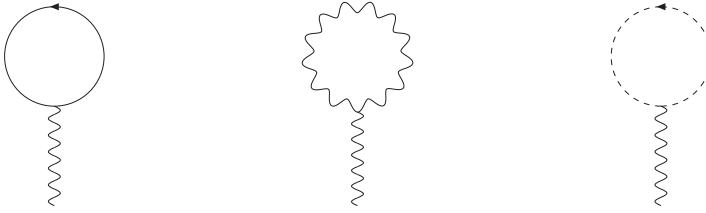
\begin{figure}[htb]
\begin{center}
\SetScale{.75}
\begin{picture}(50,100)(25,0)
\ArrowArc(50,75)(25,-90,270)
\Photon(50,0)(50,50)36
\end{picture}
\hspace{2cm}
\begin{picture}(50,100)(25,0)
\PhotonArc(50,75)(25,14,374){3}{13}
\Photon(50,47)(50,0)36
\end{picture}
\hspace{2cm}
\begin{picture}(50,100)(25,0)
\DashArrowArc(50,75)(25,-90,270)3
\Photon(50,0)(50,50)36
\end{picture}
\end{center}
\caption{One loop tadpole diagrams}
\label{tadpole}
\end{figure}

The diagrams appearing in Fig.~\ref{tadpole}
contribute to the renormalization of the first term in Eq.~(\ref{Fterm}),
the dimension two operator $\partial_iA_j^a$, as well as to the renormalization of the dimension four operator $\partial_k \partial_l \partial_i A_j^\alpha$ contained in the last term of Eq.~(\ref{renormL}). 
In the first diagram of Fig.~\ref{tadpole} 
six dimensional fermions circulate. 
The contribution of a chiral fermion $\Psi_{\pm}$ turns out to be
\begin{equation}
ig~\text{tr}\left( \lambda_R T_R^B \right)\epsilon_{ij} m^j 
\int\frac{d^4q}{(2\pi)^4}~
\frac{1}{q^2-\vec m^2/2}
,\qquad m_5,m_6\text{ even},
\label{tadfer}
\end{equation}
where the external leg corresponds to the 4D scalar $A_i^B$ (we have
defined $\epsilon_{56}=-\epsilon_{65}=+1$). It leads to the brane
terms~\footnote{We have confirmed explicitly that the terms $f^{\a\hat
a\hat b}A^{\hat a}_iA_j^{\hat b}$ in (\ref{Fterm}) receive the same
renormalization ${\cal Z}_\a^{ij}$ at one loop as the tadpole.}
\begin{equation}
\left( {\cal
Z}_\alpha^{ij}F^\alpha_{ij}+\mathcal{Z}_\alpha^{klij}D^{\alpha A}_k
D^{AB}_l F^B_{ij}\right) \left[\delta(y_5)+\delta(y_5-\pi) \right]
\left[\delta(y_6)+\delta(y_6-\pi) \right],
\end{equation}
where $\alpha$ runs over the different $U(1)$ factors of $\cal H$ and
${\cal Z}_\alpha^{ij}$ and $\mathcal{Z}_\alpha^{klij}$ are given by
\begin{equation}
{\cal Z}_\alpha^{ij}=\epsilon^{ij}\,\frac{g}{32\pi^2}\,
\zeta^\alpha\,\Lambda^2,\quad \zeta^\alpha=\text{tr}\left( \lambda_R
T_R^\alpha \right),
\end{equation}
\begin{equation}
{\cal Z}_\alpha^{klij}=\delta^{kl}\,\epsilon^{ij}\,\frac{g}{32\pi^2}\,
\zeta^\alpha\,\log\frac{\Lambda}{\mu},
\end{equation}
where $\Lambda$ is the ultraviolet and $\mu$ the infrared cut-off.

A further comment concerns the gauge contribution to the tadpole. At
one loop it is given by the second (contribution from gauge fields
$A_M$) and third (contribution from ghosts $c$) diagrams in
Fig.~\ref{tadpole}. Each one is proportional to the corresponding
trace
\begin{equation} 
{\rm tr}\left( \lambda_{\rm Adj} T_{\rm Adj}^\alpha
\right)=\eta^A\delta^{AB}f^{\a AB}=0 
\label{gaugetrace}
\end{equation}
and thus vanish by the asymmetry of the structure
constants. Note that this is a generic feature of real
representations. 

We have also computed the one loop contribution to the terms
$(F_{ij}^a)^2$ in (\ref{renormL}) and we found the 
logarithmic divergence we anticipated:
\begin{equation}
-\frac{1}{4}F^a_{ij} {\cal Z}_{ab}^{ijkl} F^{b}_{kl}
\left[\delta(y_5)+\delta(y_5-\pi) \right]
\left[\delta(y_6)+\delta(y_6-\pi) \right],
\end{equation}
where
\begin{equation}
{\cal Z}_{ab}^{ijkl}=\frac{1}{2}\left(\delta^{ik}\delta^{jl}-
\delta^{jk}\delta^{il}\right)
\frac{g^2}{2\pi^2}\
\left( C_2({\cal H}_a)-\frac{1}{2}C_2({\cal G}) \right)
\delta_{ab}\ \log\frac{\Lambda}{\mu}.
\label{zeta}
\end{equation}
Here, $C_2({\cal H}_a)$ is by
definition the Casimir of the group factor in ${\cal H}$ to which the
generator $T^a$ belongs (we define it to be zero for $U(1)$
factors). One expects a corresponding logarithmic contribution from
the fermion sector.

Since we have seen that, in all cases, the one loop contribution to
the renormalization of (\ref{Fterm}) from six dimensional gauge
bosons and fermions is proportional to $\tr(\lambda T^\a)$, where
$\lambda$ represents the parity action on the corresponding
representation, two main issues can be addressed. The first issue
concerns the vanishing of the tadpole from the gauge sector (or in
general from real representations) at higher orders in perturbation
theory. In order to answer this question we have computed the two loop
contribution to the tadpole.
We have verified (see appendix) that the contribution of real
representations to the tadpole vanish at higher loop order and we can
expect that it vanishes at all orders in perturbation theory although
we do not have an explicit proof beyond two loops.

The second issue concerns the possible relation between the tadpole
and the generation of the four dimensional anomalies on the brane by
(chiral) fermions in the bulk
\cite{Arkani-Hamed:2001is,Asaka:2002}. In fact we have seen that given
a collection of six dimensional chiral fermions $\Psi_\varepsilon$,
where $\varepsilon=\pm$, the generated tadpole for a given $U(1)$
factor is
\begin{equation}
\zeta^\a = \sum_{\Psi_\varepsilon} \tr(\lambda_{\Psi_\varepsilon} 
T^\a_{\Psi_\varepsilon}),
\label{totalz}
\end{equation}
where $\varepsilon=\pm$ is the six dimensional chirality of the field
$\Psi_\varepsilon$.  On the other hand, the four dimensional anomaly
on the brane is generated by bulk triangular loop diagrams where
chiral fermions $\Psi_\varepsilon$ circulate in the loop, while gauge
bosons and/or gravitons are external legs. In particular the mixed
$U(1)$$-$gravitational anomaly on the brane is easily seen to be
proportional to
\begin{equation} 
\mathcal{A}^\a = \sum_{\Psi_\varepsilon}
\varepsilon\ \tr(\lambda_{\Psi_\varepsilon} T^\a_{\Psi_\varepsilon}).
\label{anomaly}
\end{equation} 
Notice that different chiralities contribute with the same sign to the
tadpole $\mathcal{Z}^\a$ while they contribute with different signs to
the anomaly $\mathcal{A}^\a$.  Let us also note that with the choice
$\mathcal{P}'_{\frac{1}{2}}$ the $\varepsilon$ would move from
Eq.~(\ref{anomaly}) to Eq.~(\ref{totalz}). Keeping the physics
constant requires however to make the change
$\lambda\rightarrow-\lambda$ for the negative chirality fermions (see
footnote \ref{bla}), which is consistent with
Eqs.~(\ref{totalz}) and (\ref{anomaly}).

By looking at (\ref{totalz}) and (\ref{anomaly}) one may conclude that
imposing $\zeta^{\a}={{\cal A}^{\a}}=0$ results in the tadpole
cancellation to be equivalent to the $U(1)-$gravitational anomaly
cancellation in the positive and negative chirality sectors
separately.  However, in models originating from string theories
anomalies can be cancelled by a generalized Green-Schwarz
mechanism. In those cases the cancellation of the anomaly ${{\cal
A}^{\a}}$ in Eq.~(\ref{anomaly}) is no longer a necessary condition
and therefore the tadpole cancellation as given by Eq.~(\ref{totalz})
remains as the only constraint in the model.  We will illustrate the
above ideas with the six-dimensional model of Ref.~\cite{ABQ}
compactified on $T^2/\mathbb{Z}_2$ with gauge group
$\mathcal{G}=SU(3)_c\times SU(3)_w\times U(1)_{\mathcal{Q}_3} \times
U(1)_{\mathcal{Q}_2}$ broken by the orbifold boundary conditions to
$\mathcal{H}=SU(3)_c\times SU(2)_w\times U(1)_{\mathcal{Q}_1}\times
U(1)_{\mathcal{Q}_3} \times U(1)_{\mathcal{Q}_2}$. Fermions are in
representations $L_f=({\bf 1},{\bf 3})^+_{(0,1)}$, $U_f=({\bf 3},{\bf
1})^+_{(1,0)}$, $Q_f=({\bf 3},{\bf 3})^{\varepsilon_f}_{(1,1)}$ where
$f=1,2,3$, $\varepsilon_{1,2}=-$, $\varepsilon_{3}=+$, and the
notation $({\bf r_3},{\bf r_2})^\varepsilon_{(q_3,q_2)}$ represents a
six dimensional Weyl fermion with chirality $\varepsilon$ in the
representation ${\bf r_3}$ and ${\bf r_2}$ of $SU(3)_c$ and $SU(3)_w$,
respectively, and $U(1)$ charges $q_3$ and $q_2$ under the generators
$\mathcal{Q}_3$ and $\mathcal{Q}_2$. Orbifold compactification breaks
$SU(3)_w\to SU(2)_w\times U(1)_{\mathcal{Q}_1}$, where $\mathcal{Q}_1=
diag(1,1,-2)$ and for $SU(3)_w$ triplets the matrix satisfying
(\ref{auto2}) is $\lambda=diag(1,1,-1)$. The Standard Model
hypercharge is related to $\mathcal{Q}_i$ by $Y=\mathcal{Q}_1/6-2
\mathcal{Q}_2/3+2\mathcal{Q}_3/3$ and the fields are decomposed under
$SU(2)_w\times U(1)_{\mathcal{Q}_1}$ as
\begin{equation}
L_L=\left(
\begin{array}{c}
\ell_L\\
\widetilde{e}_L
\end{array}
\right),\
L_R=\left(
\begin{array}{c}
\widetilde{\ell}_R\\
{e}_R
\end{array}
\right),\
Q_L=\left(
\begin{array}{c}
q_L\\
\widetilde{d}_L
\end{array}
\right),\
Q_R=\left(
\begin{array}{c}
\widetilde{q}_R\\
{d}_R
\end{array}
\right),\
U_L=\widetilde{u}_L,\ U_R=u_R
\label{fields}
\end{equation}
where untilded (tilded) fields are (mirrors of) Standard Model fields.

Given (\ref{fields}) the parity properties of fields is given by:
$\lambda_L=\lambda$, $\lambda_{Q_3}=\lambda$,
$\lambda_{Q_{1,2}}=-\lambda$, $\lambda_U=-1$. Thus their contribution to $\zeta^1$ cancels while $\mathcal{A}^1=12(1+N_c)$ since, as stressed in Ref.~\cite{ABQ}, $\mathcal{Q}_1$ is anomalous.

An alternative way to the Green-Schwarz mechanism, that can be used to
cancel the (bulk-induced) brane anomalies in Eq.~(\ref{anomaly}), is
by means of chiral fermions localized on the brane. Localized fermions
do not possess tree level couplings with $A_i^A$, or their
derivatives, and thus they provide no one-loop contribution to the
tadpole. Moreover their unique two loop contribution, given by the
third diagram of Fig.~\ref{twoloops} where the dashed (ghost) line is
replaced by a localized chiral fermion, vanishes as can be easily
checked.

We want to conclude this letter by stressing the fact that the
conditions for tadpole cancellation on the brane do not coincide with
those required from bulk-induced anomaly cancellation. As such the
tadpole is not expected to be (as the anomaly) a purely one loop
effect and in a general theory with fermions we expect a tadpole
generation at least at the two loop level. However for theories with
low (TeV) cut-off scale the latter will provide a mild (tiny)
dependence on the cut-off that should not disturb the stability of the
low energy effective theory.

\section*{\sc\Large Acknowledgments} 
We thank L.~Covi, C.~Cs\'aki, C.~Grojean and E.~Kiritsis for
discussions, and S.~Groot Nibbelink for some useful remarks.  We also
thank C.~Cs\'aki, C.~Grojean and H.~Murayama for sharing their recent
work~\cite{CGM} with us prior to publication.  The work of GG was
supported by the DAAD.
%
\newpage
\appendix
\@addtoreset{equation}{section}   
\makeatother   
\renewcommand{\theequation}{\thesection.\arabic{equation}}  

\vspace{1cm}
\noindent
{\sc\Large Appendix}
\vspace{-.5cm}
\section{\Large\sc Two loop contribution to the tadpole}

\begin{figure}[htb]
\begin{center}
\SetScale{.75}
\begin{picture}(50,135)(0,0)
\PhotonArc(25,75)(25,194,554){3}{13}
\PhotonArc(25,121)(15,-11,349){3}8
\Photon(25,0)(25,52)3{4.5}
\end{picture}
\hspace{1cm}
\begin{picture}(50,100)(0,0)
\PhotonArc(25,75)(25,14,374){3}{13}
\Photon(50,75)(0,75)3{4.5}
\Photon(25,47)(25,0)3{4.5}
\end{picture}
\hspace{1cm}
\begin{picture}(50,100)(0,0)
\PhotonArc(25,75)(25,14,374){3}{13}
\SetColor{White}
\BCirc(25,95){10}
\SetColor{Black}
\DashArrowArc(25,95)(10,-90,270)3
\Photon(25,47)(25,0)3{4.5}
\end{picture}
\hspace{1cm}
\begin{picture}(50,100)(0,0)
\DashArrowArc(25,75)(25,-90,270)3
\Photon(50,75)(0,75)3{4.5}
\Photon(25,0)(25,50)3{4.5}
\end{picture}
\hspace{1cm}
\begin{picture}(50,100)(0,0)
\PhotonArc(25,75)(25,104,464){3}{13}
\Photon(25,100)(25,0)39
\end{picture}
\end{center}
\caption{Two loop tadpole diagrams from the gauge sector}
\label{twoloops}
\end{figure}
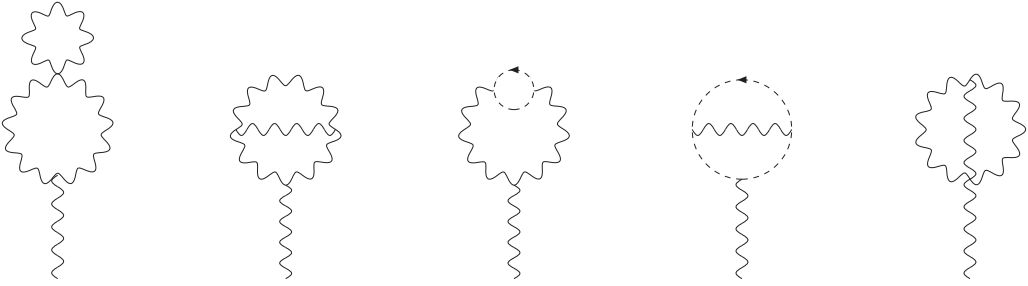

The diagrams contributing to the tadpole at two loops are given in
Fig.~\ref{twoloops}.  Note that fermionic diagrams are obtained by
just replacing in the diagrams of (\ref{twoloops}) ghost propagators
by fermion propagators.
To judge whether or not there are contributions to the tadpole, it is
sufficient to examine the gauge index structure.  Recall that the
gauge and ghost propagators have the general structure (see
Eq.~(\ref{propagator}))
\begin{equation}
\delta_+\delta_{AB}+\delta_-\Lambda_{AB}=(\delta_{+}\!+\delta_-\eta_A)
\delta_{AB}
\end{equation}
while for fermions in the representation $R$ we have
\begin{equation}
\delta_+\delta_{ij}+\delta_-(\lambda_R)_{ij}.
\end{equation}
Here $\delta_+$ symbolizes extra dimensional momentum conservation and
$\delta_-$ extra dimensional momentum flip.  The first four diagrams
are reducible in the sense that they just correspond to wave function
renormalization insertions of the gauge or ghost propagators.  Using
the contraction identities~\footnote{The tensor in Eq.~(\ref{oneeta})
vanishes for $B=\hat b$. It has already been encountered in
Eq.~(\ref{zeta}). }
\begin{eqnarray}
f^{BDE}f^{CDE}&=&C_2({\cal G})\delta^{BC}\\
f^{BDE}f^{CDE}\eta^D&=&\left(C_2({\cal H}_B)- 
\frac{1}{2}C_2({\cal G})\right)(\eta_B+1)\delta^{BC}\label{oneeta}\\
f^{BDE}f^{CDE}\eta^D\eta^E&=&C_2({\cal G})\eta^B\delta^{BC}
\end{eqnarray}
one can verify immediately that all these insertions are matrices
${\cal Z}_{BC}$ which are symmetric in $BC$. These are then to be
contracted with $f^{ABC}$, giving zero.  In a similar way it can be
seen that the last diagram does not contribute either.

Finally, fermions contribute at two loops unless they transform in a
real representation, $T_R^T=-T_R$. The corresponding contraction
identities read:
\begin{eqnarray}
\text{tr}~(T_R^BT_R^{C})&=&C_R\delta^{BC}\\
\text{tr}~(T_R^B\lambda_R T_R^{C}\lambda_R)&=&C_R\eta^B\delta^{BC},
\end{eqnarray}
and in addition $\text{tr}~(T_R^B T_R^{C}\lambda_R)=\text{tr}~(T_R^{C}
T_R^{B}\lambda_R)$ if $R$ is real. We conclude that in this case the
third diagram (with the ghost replaced by a fermion) is zero, while
for general $R$ there will be a contribution from the antisymmetric
part of $\text{tr}~(T_R^B T_R^{C}\lambda_R)$\footnote{It is easy to
verify that this corresponds to a brane term.}.  Finally, the fourth
diagram can be seen to give terms proportional to the four tensors
(the sum over $B$ is understood)
\begin{equation}
\text{tr}~(T^a T^B T^B),\quad \eta_B\text{tr}~(T^a T^B T^B),
\quad \text{tr}~(T^a \lambda T^B T^B),
\quad \eta_B\text{tr}~(T^a \lambda T^B T^B).
\end{equation}
The Casimirs $T^BT^B$ and $\eta_BT^BT^B$ are symmetric matrices which
commute with $T^a$. Together with Eq.~(\ref{auto2}) this implies
that all four traces vanish for $R$ real.

\end{document}